\RequirePackage{ifpdf}
\ifpdf % We are running pdfTeX in pdf mode
\documentclass[pdftex]{sigma}
\else
\documentclass{sigma}
\fi

\begin{document}
\allowdisplaybreaks

\renewcommand{\PaperNumber}{069}

\renewcommand{\thefootnote}{$\star$}

\FirstPageHeading

\ShortArticleName{Integrable Models of Interaction of Matter with
Radiation}

\ArticleName{Integrable Models of Interaction of Matter \\ with
Radiation\footnote{This paper is a contribution 
to the Vadim Kuznetsov Memorial Issue ``Integrable Systems and Related Topics''.
The full collection is available at 
\href{http://www.emis.de/journals/SIGMA/kuznetsov.html}{http://www.emis.de/journals/SIGMA/kuznetsov.html}}}

\Author{Vladimir I. INOZEMTSEV~$^\dag$ and Natalia G.
INOZEMTSEVA~$^\ddag$} \AuthorNameForHeading{V.I. Inozemtsev and
N.G. Inozemtseva}

\Address{$^\dag$~Laboratory of Theoretical  Physics, JINR, Dubna, Russia}
\EmailD{\href{mailto:inozv@theor.jinr.ru}{inozv@theor.jinr.ru}}

\Address{$^\ddag$~Moscow Technical University, Dubna Branch, Russia}

\ArticleDates{Received July 18, 2006, in f\/inal form September
19, 2006; Published online October 13, 2006}

\Abstract{The simplif\/ied models of interaction of charged matter
with resonance modes of radiation generalizing the well-known
Jaynes--Cummings and Dicke models are considered. It is found that
these new models are integrable for arbitrary numbers of dipole
sources and resonance modes of the radiation f\/ield. The problem
of explicit diagonalisation of corresponding Hamiltonians is
discussed.}

\Keywords{integrability; radiation; Gaudin models}

\Classification{81R05; 81S99}

\section{Introduction}
In the last four decades, much attention has been paid to the
problem of describing processes of interaction of charged matter
with electromagnetic radiation at resonance \cite{Dick, Cum, Hei,
Allen}. This problem, being
 overcomplicated from the mathematical
viewpoint in the general case, admits some very attractive
simplif\/ications which allow one to construct some rather simple
models having even {\it exact} solutions. These models can be used
for quantum statistical description of real interaction processes.

The starting point for constructing such models is the quantum
Hamiltonian \cite{Hei}
\begin{gather*}
 H={1\over 2m}\left(\vec p-{e\over c}\vec A\right)^2+u(x)+H_{F},\qquad
H_{F}=\sum_{k}\omega_{k}\left(b_{k}^{+}b_{k}+{1\over 2}\right),
\end{gather*}
where $\vec p$ and $m$ are the momentum and mass of the electron
in some atom, $u(x)$ is the Coulomb potential of interaction of
the electron with nucleus, and $\vec A$ is
 the vector potential of the secondly quantized electromagnetic f\/ield
which is given by the formula
\begin{gather*}
\vec A=\sum_{k}\left({{2\pi c^2}\over{\omega_{k}}}\right)^{1/2}
\vec e_{k}\left(b_{k}^{+}e^{-i\vec k \vec x}+b_{k}e^{i\vec k \vec
x}\right), \qquad [b_{k},b_{k'}^{+}]=\delta_{kk'},
\end{gather*}
where $\vec e_{k}$ are polarization vectors and $b_{k}$,
$b_{k}^{+}$ are the usual operators of annihilation and creation
of the mode with wave vector $\vec k$ and the energy $\omega_{k}$.
There are several steps of simplif\/ication of this general
Hamiltonian. First, one neglects the term which is nonlinear in
the vector potential, ${{e^2}\over{2mc^2}}\vec A^2$. Second, one
considers only transitions between higher levels of an atom,
\[
 {1\over 2m}\vec p\,{}^2 +u(x)\rightarrow \omega_{0}\sigma^{z}+
 {\rm const},
 \] where $\sigma^{z}$ is the Pauli matrix. Third, one considers only
 small $\vec k$, $ ka\ll 1$, where $a$ is the typical size of an atom,
 $ e^{\pm i\vec k\vec x}\sim 1$.
Fourth, one uses the dipole and so-called rotating wave
approximations~\cite{Allen}, which result in
\[
-{e\over mc}\vec p \vec A\quad\rightarrow\quad
\sum_{k}g_{k}\sigma^{x} (b_{k}^{+}+b_{k})\quad\rightarrow\quad
\sum_{k}g_{k}(b_{k}^{+}\sigma^{-} +b_{k}\sigma^{+}),
\] where $\sigma^{\pm}$ are the usual Pauli matrices.
And f\/inally, one considers only the case of
 resonance: $\sum_{k}\rightarrow k_{\rm fixed}$. By all these steps one can
 write
down very simple Jaynes--Cummings Hamiltonian~\cite{Cum} which is
linear in the operators of creation and annihilation of {\it one}
resonant mode,
\begin{gather}
\label{eq1}
 H_{JC}=\omega_{0}b^{+}b +{1\over 2}\omega_{0}\sigma^{z}
+g(\sigma^{+}b+\sigma^{-}b^{+}).
\end{gather}
The solution of eigenequation for the operator \eqref{eq1} is
almost trivial, but it allows one to describetime evolution of the
corresponding wave function in great detail, with transparent
applications to physics of one-mode laser. This extremely simple
model admits almost evident generalisations in two directions.

\section[The simplest generalisations of the JC model:\\
 the $n$-level atom and Dicke model]{The simplest generalisations of the JC model:\\
 the $\boldsymbol{n}$-level atom and Dicke model}

The f\/irst of these generalisations is the model of one $n$-level
atom with $(n-1)$ modes of resonant radiation,
\begin{gather}
\label{eq2} H_{JC}^{(n)}=\sum_{\alpha=1}^{n-1}\omega_{\alpha}\hat
N_{\alpha} +\Delta\hat
R_{nn}+\sum_{\alpha=1}^{n-1}g_{\alpha}\big(b_{\alpha}\hat
R_{n\alpha} +b_{\alpha}^{+}\hat R_{\alpha n}\big),
\end{gather}
where
\[
\hat N_{\alpha}=b_{\alpha}^{+}b_{\alpha}-\hat
R_{\alpha\alpha},\qquad \big(\hat
R_{\alpha\beta}\big)_{\gamma\delta}=\delta_{\alpha\gamma}
\delta_{\beta\delta},
\]
the sets $\{\omega_{\alpha}\}$, $\{g_{\alpha}\}$ are energies of
the modes and their coupling constants, and $\Delta$ is so-called
detuning parameter \cite{Li, Abdel}. The extended physical
motivation for this kind of the generalization of the
Jaynes--Cummings model can be found in \cite{Abdel}. There are
evident commutation relations
\[
\big[\hat N_{\alpha},\hat N_{\beta}\big]=0, \qquad
\big[H_{JC}^{(n)}, \hat N_{\alpha}\big]=0,
\]
which make the problem of diagonalisation of \eqref{eq2} at
$n\geq3$ quite easy. The second way of generalisation of the
Jaynes--Cummings model consists in considering an arbitrary number
$N$ of two-level sources of radiation,
\begin{gather}
\label{eq3} H_{D}=\omega b^{+}b +\sum_{j=1}^{N}
\big[\omega_{j}\sigma_{z}^{(j)}+
\lambda\big(\sigma_{+}^{(j)}b+\sigma_{-}^{(j)}b^{+}\big)\big].
\end{gather}
 The solution of this so-called Dicke model which has been proposed many
years ago \cite{Dick} is far from being trivial. It turned out
that the problem with the Hamiltonian~\eqref{eq3} has some hidden
{\it dynamical} symmetry which allows one to construct some
nontrivial operators commuting with~\eqref{eq3}.

What happens if one will try to use both the ways of
generalisation of the Jaynes--Cummings model, i.e.\ consider the
Hamiltonian of  the problem of an arbitrary number $N$ of
multilevel sources interacting with an arbitrary number $n-1$ of
resonant modes of radiation,
\begin{gather}
\label{eq4} H^{(N,n)}=\sum_{\alpha=1}^{n-1}\omega_{\alpha}\hat
N_{\alpha} +\sum_{k=1}^{N}\Delta_{k}\hat
R_{nn}^{(k)}+\sum_{\alpha=1}^{n-1} \sum_{k=1}^{N}
g_{\alpha}^{(k)}\big[b_{\alpha}\hat
R_{n\alpha}^{(k)}+b_{\alpha}^{+} \hat R_{\alpha n}^{(k)}\big],
\end{gather}
where $ N\geq 2$, $n\geq 3$ (the detuning parameters $\{\Delta\}$
and coupling constants $\{g_{\alpha}^{(k)}\} $ are assumed to be
arbitrary)? The answer is not clear till now. In the following two
sections we will argue that there is some simplif\/ication of the
spectral problem def\/ined by \eqref{eq4} in the case  of equality
of all the modules of the coupling constants $\{g\}$: these
constants should not depend on $k$ and their dependence on
$\alpha$ is given by the formula $g_{\alpha}^{(k)}\propto
\varepsilon_{\alpha}$, $\varepsilon_{\alpha}=\pm 1$.
 Namely, we will construct in this case
explicitly some set of $N-1$ nontrivial operators commuting with
\eqref{eq4}; their existence allows one to call the model
\eqref{eq4} {\it quantum integrable} under the above choice of the
coupling constants.

\section{The models of Gaudin type}

In this section, let us consider some more general operators
introduced f\/irst by Gaudin~\cite{Gau1},
\begin{gather}
\label{eq5} {\cal
H}_{j}^{(N)}=\sum_{\alpha,\beta=1}^{n}\sum_{s\neq j}^{N+1}
f_{js}^{\alpha\beta}{\hat T}_{\alpha\beta}^{j}{\hat
T}_{\beta\alpha}^{s}, \qquad j=1,\dots,N+1,
\end{gather}
where $\left\{{\hat T}_{\alpha\beta}^j\right\}$ are operators of
absolutely {\it arbitrary}
 representations of
$SL(n)$, obeying the commutation relations
\begin{gather}\label{eq6}
\big[{\hat T}_{\alpha\beta}^{j}, {\hat T}_{\lambda\mu}^{s}\big]=
\delta_{js}\big({\hat T}_{\alpha\mu}^{j}\delta_{\beta\lambda}-
{\hat T}_{\lambda\beta}^{j}\delta_{\alpha\mu}\big),
\end{gather}
and $\big\{f_{js}^{\alpha\beta}\big\}$ are arbitrary sets of
numbers, obeying the evident restriction
$f_{js}^{\alpha\beta}=f_{js}^{\beta\alpha}.$ It is easy to see
that operators
\[
{\hat N}_{\alpha}=\sum_{j=1}^{N+1}{\hat T}_{\alpha\alpha}^{j}
\]
trivially commute with \eqref{eq5},
\[
\big[{\cal H}_{j}^{(N)},{\hat N}_{\alpha}\big]=0.
\]
Now let us pose the problem: how to f\/ind such the sets of
numbers $\{f\}$ that
\begin{gather}
\label{eq6+} \big[{\cal H}_{j}^{(N)}, {\cal H}_{k}^{(N)}\big]=0
\end{gather}
for all indices $\{j,k\}$.
 It is easy to see that in this case the ``Hamiltonian''
\begin{gather}\label{eq7}
H_{n}^{(N)}=\sum_{j=1}^{N+1}\eta_{j} {\cal
H}_{j}^{(N)}+\sum_{\alpha=1}^n\omega_{\alpha}{\hat N}_{\alpha},
\end{gather}
where $\eta_{j}$ and $\omega_{alpha}$ are arbitrary constants,
commutes with all ${\cal H}_{j}^{(N)}$,
\[
\big[H_{n}^{(N)},{\cal H}_{j}^{(N)}\big]=0.
\]
In the simplest nontrivial case of $n=2$, one can always choose
$\{f\}$ such that $f_{jk}^{\alpha\beta}=-f_{kj}^{\beta\alpha},$
and the equations \eqref{eq6+} can be written in the form
\begin{gather}
\big[{\cal H}_{j}^{(N)}, {\cal H}_{k}^{(N)}\big]= -\sum_{s\neq
j,k}^{N+1}\sum_{\alpha,\beta,\lambda=1}^{n}
\big[f_{js}^{\alpha\beta}f_{sk}^{\alpha\lambda}+f_{kj}^{\alpha\lambda}
f_{js}^{\beta\lambda}+f_{jk}^{\alpha\beta}f_{ks}^{\beta\lambda}\big]\nonumber\\
\phantom{\big[{\cal H}_{j}^{(N)}, {\cal H}_{k}^{(N)}\big]=}{}
\times \big({\hat T}_{\alpha\beta}^{j} {\hat
T}_{\lambda\alpha}^{k} {\hat T}_{\beta\lambda}^{s}- {\hat
T}_{\beta\alpha}^{j} {\hat T}_{\alpha\lambda}^{k} {\hat
T}_{\lambda\beta}^{s}\big)=0, \label{eq8}
\end{gather}
which immediately gives a overdetermined set of bilinear equations
for $\{f\}$,
\begin{gather}
f_{js}^{\alpha\beta}f_{sk}^{\alpha\lambda}+f_{kj}^{\alpha\lambda}
f_{js}^{\beta\lambda}+f_{jk}^{\alpha\beta}f_{ks}^{\beta\lambda}=0.\label{eq9}
\end{gather}
It has been shown by Gaudin~\cite{Gau1, Gau2} that for $n=2$
there is some so-called anisotropic solution to~\eqref{eq9},
\begin{gather}
f_{js}^{\alpha\alpha}=\cot (u_{j}-u_{s}),\qquad
f_{js}^{\alpha\beta}={1\over{\sin(u_{j}-u_{s})}},\label{eq10}
\end{gather}
where $\left\{u_{j}\right\}$ are arbitrary numbers. Now, following
Gaudin, let us choose them as
\[
u_{k}=-\varepsilon \lambda_{k}+{\pi\over 2}+u_{N+1},\qquad
k=1,\dots, N.
\]
As $\varepsilon\to 0$, then $
f_{j,N+1}^{\alpha\alpha}\sim\varepsilon \lambda_{j}$,
$f_{j,N+1}^{\alpha\beta}\sim 1$. Let us choose $\big\{{\hat
T}_{\alpha\beta}^{N+1}\big\}$ as
 Jordan--Schwinger
realization of $SL(2)$, $ {\hat
T}_{\alpha\beta}^{N+1}=-b_{\alpha}b_{\beta}^{+}$ and consider the
limit $ N_{2}\gg 1$, where $N_{2}$ is an eigenvalue of the
operator $\hat N_{2}$.
 It is easy to see that in this limit $b$-operators can be
treated as $c$-numbers; $b_{2},b_{2}^{+}\sim\sqrt{N}_{2}$; let us
also take $ \varepsilon \sim N_{2}^{-1\over 2}$ and send
$N_{2}\to\infty$. In this limiting process,
 one easily f\/inds
\begin{gather}
\lim_{N_{2}\to\infty}H_{2}^{(N)}=\omega_{1}b_{1}^{+}b_{1}
+\sum_{j=1}^{N}\big[\lambda_{j}\big({\hat T}_{11}^{(j)}-{\hat
T}_{22}^{(j)} \big)+\big({\hat T}_{12}^{(j)}b_{1}+{\hat
T}_{21}^{(j)}b_{1}^{+}\big) \big].\label{eq11}
\end{gather}
And f\/inally, if we will choose $\big\{{\hat
T}_{\alpha\beta}^{(j)}\big\}$ as $\sigma$-representation of
$SL(2)$, we will obtain just the Dicke Hamiltonian!

This derivation of the Dicke model proposed f\/irst by Gaudin
\cite{Gau1} is not original of course. There are some recent
papers in which the connection between it and  the Gaudin models
was exploited for construction the algebraic Bethe ansatz method
for eigenvectors of the Dicke Hamiltonian and some generalisations
which might be considered as the interaction term without the
rotating wave approximation~\cite{Jurco,Duk,Kundu}.

The above limiting procedure, however, does not work in the
general case $ n\geq 3$. One also has in this case the
overdetermined system of bilinear equations
\begin{gather*}
f_{js}^{\alpha\beta}f_{sk}^{\alpha\lambda}
+f_{kj}^{\alpha\lambda}f_{js}^{\beta\lambda}+f_{jk}^{\alpha\beta}
f_{ks}^{\beta\lambda}=0,
\end{gather*}
but, if $n\geq 3$, there is a possibility of the choice of indices
 $\alpha\neq\beta\neq\lambda$ which excludes all ``anisotropic'' solutions
of the type \eqref{eq10}. There are only ``isotropic'' solutions
of the type
 $f_{js}^{\alpha\alpha}=f_{js}^{\alpha\beta}={1\over
{\nu_{j}-\nu_{s}}},$ and hence there are no parameters for
$N_{n}\to\infty$.
 Hence it seems that the Gaudin approach does not lead to any integrable
system of the type \eqref{eq4}.

Let us prove absence of ``anisotropic'' solutions to \eqref{eq9}
as $N=2$, $n\geq 3$. In this simplest nontrivial case the set
$\{f\}$ can be parametrized as
\begin{gather}
f_{js}^{\alpha\beta}=\sum_{p=1}^3 \varepsilon_{jsp}A_{p}^
{\alpha\beta},\qquad f_{sk}^{\alpha\lambda}=\sum_{p=1}^3
\varepsilon_{skp} A_{p}^{\alpha\lambda}, \qquad
f_{js}^{\beta\lambda}=\sum_{p=1}^3
\varepsilon_{jsp}A_{p}^{\beta\lambda}.\label{eq12}
\end{gather}
As
 $j\neq k\neq s,$ \eqref{eq9} reads:
\[
A_{k}^{\alpha\beta}A_{j}^{\alpha\lambda}+
A_{s}^{\alpha\lambda}A_{k}^{\beta\lambda}+A_{s}^{\alpha\beta}A_{j}^{\beta
\lambda}=0,
\]
with general solution
\begin{gather*}
A_{1}^{\alpha\lambda}={A\over{\sin \nu_{1}}}, \qquad
A_{2}^{\alpha\lambda}={A\over{\sin\nu_{2}}},\qquad
A_{3}^{\alpha\lambda}
=-{A\over{\sin(\nu_{1}+\nu_{2})}},\\
A_{1}^{\alpha\alpha}=A\cot\nu_{1},\qquad A_{2}^{\alpha\alpha}=
A\cot\nu_{2},\qquad A_{3}^{\alpha\alpha}=-A\cot(\nu_{1}+\nu_{2}),
\end{gather*}
where $ A$, $\nu_{1}$, $\nu_{2}$ are some parameters. Now, if one
takes $\alpha\neq\beta\neq\lambda$, one gets from \eqref{eq9}
\[
 -{{A^2}\over{2\cos{{nu_{1}}\over2}
\cos{{\nu_{2}}\over2}\cos{{\nu_{1}+\nu_{2}}\over2}}}=0\quad
\to\quad A=0.
\]
 Surprisingly enough, one can construct integrable models of Gaudin type
at $n\geq 3$ by using only ``isotropic'' solution of \eqref{eq9}.
It will be done in the next section.

\section[New integrable model with $N$ sources and $n-1$ modes]{New
integrable model with $\boldsymbol{N}$ sources and
$\boldsymbol{n-1}$ modes}

The receipt of constructing the integrable model of the type
\eqref{eq4} is quite unusual. Let us choose the above $\cal
H$-operators as obeying
 the equality
\begin{gather}\label{eq13}
\big[{\cal H}_{j}^{(N)},{\cal H}_{k}^{(N)}\big]= {\hat
S}_{jk}\big(\big\{{\hat T}_{\alpha\beta}^{s}\big\},\{x\}\big),
\end{gather}
where  $x$ is some parameter and ${\hat S}_ {jk}$ have the
following property: as $x\to\infty,$ there are
 only zero eigenvalues of them in a subspace in
 which
 $\big\{{\hat T}_{\alpha\beta}^{s}\big\}$ act nontrivially.
Then we can introduce new operators $\big(\lim\limits_{x\to\infty}
{\cal H}_{j}^{(N)}\big)$ in this subspace which will commute,
 and $H_{n}^{N+1}=-\sum\limits_{j=1}^{N}{\cal H}_{j}^{(N)}$ is
some nontrivial Hamiltonian. By direct calculation of the
commutator of the operators \eqref{eq5}, one f\/inds
\begin{gather}
{\hat S}_{jk}\big(\big\{{\hat T}_{\alpha\beta}^{s}\big\}
,\{x\}\big\}=-\sum_{s\neq
j,k}^{N}\sum_{\alpha,\beta,\lambda=1}^{n}
\big(f_{js}^{\alpha\beta}f_{sk}^{\alpha\lambda}
+f_{kj}^{\alpha\lambda}f_{js}^{\beta\lambda}
+f_{jk}^{\alpha\beta}f_{ks}^{\beta\lambda}\big)\nonumber\\
\qquad{}\times \big( {\hat T}_{\alpha\beta}^{j}{\hat
T}_{\lambda\alpha}^{k} {\hat T}_{\beta\lambda}^{s}- {\hat
T}_{\beta\lambda}^{j}{\hat T}_{\alpha\lambda}^{k} {\hat
T}_{\lambda\beta}^{s} \big)-
\sum_{\alpha,\beta,\lambda=1}^{n}\big(
f_{j,N+1}^{\alpha\beta}f_{N+1,k}^{\alpha\lambda}+
f_{kj}^{\alpha\lambda}f_{j,N+1}^{\beta\lambda}+f_{jk}^{\alpha\beta}f_{k,N+1}^{\beta\lambda}\big)\nonumber\\
\qquad{}\times \big( {\hat T}_{\alpha\beta}^{j}{\hat
T}_{\lambda\alpha}^{k} {\hat T}_{\beta\lambda}^{N+1}- {\hat
T}_{\beta\lambda}^{j}{\hat T}_{\alpha\lambda}^{k} {\hat
T}_{\lambda\beta}^{N+1} \big).\label{eq14}
\end{gather}
Let us choose  $f_{jk}^{\alpha\beta}=A(\nu_{j}-\nu_{k})^{-1}$,
$1\leq j,k\leq N$, i.e.\ use the
 isotropic Gaudin solution at $j,s,k\leq N$.
 Then the double sum vanishes and we are left with
\begin{gather}
 {\hat S}_{jk}\big(\big\{{\hat T}_{\alpha\beta}^{s}\big\},
\{x\}\big)=-\sum_{\alpha,\beta,\lambda=1}^{n}\big[
f_{j,N+1}^{\alpha\beta}f_{N+1,k}^{\alpha\lambda}+
{A\over{\nu_{j}-\nu_{k}}}
\big(f_{k,N+1}^{\beta\lambda}-f_{j,N+1}^{\beta\lambda}\big)\big]\nonumber\\
\phantom{ {\hat S}_{jk}\big(\big\{{\hat
T}_{\alpha\beta}^{s}\big\},\{x\}\big)=}{} \times \big( {\hat
T}_{\alpha\beta}^{j}{\hat T}_{\lambda\alpha}^{k} {\hat
T}_{\beta\lambda}^{N+1}- {\hat T}_{\beta\lambda}^{j}{\hat
T}_{\alpha\lambda}^{k} {\hat T}_{\lambda\beta}^{N+1}
\big).\label{eq15}
\end{gather}
{\samepage There are no other restrictions to the numbers
$\big\{f_{j,N+1}^{\alpha\beta}\big\}$ except symmetry. Hence we
have $ {{Nn(n+1)}\over2}$ free parameters at given $\{A,
\{\nu_{j}\}\}$ in the relation \eqref{eq15}. Of course, there is
trivial case if  all $f_{j,N+1}^{\alpha\beta}=0.$ Fortunately, one
can construct the  nontrivial ansatz:
\[
f_{j,N+1}^{\alpha\beta}=\big(\delta_{\beta n} f_{j,N+1}^{\alpha
n}+\delta_{\alpha n}f_{j,N+1}^{\beta n}\big)
\left(1-\delta_{\alpha n}\delta_{\beta n}\right)f_{j,N+1}^{nn},
\]
i.e.\ $f_{j,N+1}^{\alpha\beta}=0$ unless $\alpha\neq n$ or
$\beta\neq n$.}

The commutator $\big[{\cal H}_{j}^{N},{\cal H}_{k}^{N}\big]$
\eqref{eq15} is still too complicated. Let us make the next
assumption:
\begin{gather}
f_{j,N+1}^{\lambda n}=\varepsilon_{\lambda}\sqrt{AF}, \qquad
f_{j,N+1}^{nn}=-F\nu_{j}, \qquad
\vert\varepsilon_{\lambda}\vert=1,\label{eq16}
\end{gather}
where $\left\{\nu_{j}\right\}$ are arbitrary numbers. Then
\eqref{eq15} can be recast in the form
\begin{gather}
\big[{\cal H}_{j}^{(N)},{\cal H}_{k}^{(N)}\big]=
\sum_{\beta,\lambda=1}^{n-1}AF\varepsilon_{\beta}\varepsilon_{\lambda}\big(
{\hat T}_{n\beta}^{j} {\hat T}_{\lambda n}^{k} {\hat
T}_{\beta\lambda}^{N+1}- {\hat T}_{\beta n}^{j} {\hat
T}_{n\lambda}^{k}
{\hat T}_{\lambda\beta}^{N+1}\big)\nonumber\\
\phantom{\big[{\cal H}_{j}^{(N)},{\cal
H}_{k}^{(N)}\big]=}{}-\sum_{\lambda=1}^{n-1}\sqrt{AF}
\big\{F\nu_{j}\varepsilon_{\lambda}\big({\hat T}_{nn}^{j} {\hat
T}_{\lambda n}^{k} {\hat T}_{n\lambda}^{N+1}- {\hat T}_{nn}^{j}
{\hat T}_{n\lambda}^{k}
{\hat T}_{\lambda n}^{N+1}\big)\nonumber\\
\phantom{\big[{\cal H}_{j}^{(N)},{\cal
H}_{k}^{(N)}\big]=}{}-F\nu_{k}\varepsilon_{\lambda}\big( {\hat
T}_{n\lambda}^{j} {\hat T}_{nn}^{k} {\hat T}_{\lambda n}^{N+1}-
{\hat T}_{\lambda n}^{j} {\hat T}_{nn}^{k} {\hat
T}_{n\lambda}^{N+1}\big)\big\}.\label{eq17}
\end{gather}
Let now choose
\[
{\hat T}_{\beta\lambda}^{N+1} =-b_{\beta}b_{\lambda}^{+}
\]
as Jordan--Schwinger representation of $SL(n)$). Hence the basis
of the whole Hilbert space of the problem consists of the direct
products of the basis vectors of the spaces in which ${\hat
T}_{\beta\lambda}^{j}$, $1\leq j\leq N$ act, and various basic
vectors of above Jordan--Schwinger representation,
$\big\{\prod_{\beta, \{l\}}
\big(b_{\beta}^{+}\big)^{l_{\beta}}\vert 0\rangle\big\}$,
 $\{l\} \in {\mathbb N}$.
 Consider the action of the right-hand side of (17) on the subspace spanned by
the vectors
 $\varphi_{nn}^{(L)}$ with $l_{n}=L$, $ L\gg 1$
such that
\begin{gather*}
{\hat N}_{n}\varphi_{nn}^{(L)}\sim L^2\varphi_{nn}^{(L)}, \qquad
{\hat N}_{n}\varphi_{nn}^{(L)}=-\left( {\hat T}_{nn}^{N+1}+
\sum_{j=1}^{N}{\hat
T}_{nn}^{j}\right)=b_{n}b_{n}^{+}-\sum_{j=1}^{N} {\hat
T}_{nn}^{j}.
\end{gather*}
 Then
\[
{\hat T}_{nn}^{N+1}\sim-L^2,\qquad {\hat
T}_{n\alpha}^{N+1}\sim-Lb_{\alpha}^{+},\qquad {\hat T}_{\alpha
n}^{N+1}\sim-Lb_{\alpha},
 \qquad 1\leq\alpha\leq n-1.
 \]
 If
 \[
 F={{A\Delta}\over{L^2}},\qquad f_{k,N+1}^{nn}=-{{A\Delta
\nu_{k}}\over{L^2}}, \qquad f_{k,N+1}^{\lambda
n}={{\varepsilon_{\lambda} A\sqrt{\Delta}}\over L},
\]
 then
 \[
 \lim_{L\to\infty}\left[
{\cal H}_{j}^{(N)},{\cal H}_{k}^{(N)}\right]\varphi_{nn}^{(L)}=0
\]
 and
 \[
 \tilde{\cal H}_{j}^{(N)}=\lim_{L\to\infty}
{\cal H}_{j}^{(N)}
\]
 commute on this subspace!
The explicit form of $\cal H$-operators reads
\begin{gather}
\tilde{\cal H}_{j}^{(N)}=A\sum_{s\neq j}^{N}\sum_{\alpha,\beta
=1}^{n}{\hat T}_{\alpha\beta}^{j}{\hat
T}_{\beta\alpha}^{s}(\nu_{j}-\nu_{s})
^{-1} +{\hat w}_{j},\nonumber\\
{\hat w}_{j}=A\left[\sum_{\lambda=1}^{n-1}\varepsilon_{\lambda}
\sqrt{\Delta}\left(b_{\lambda}{\hat
T}_{n\lambda}^{j}+b_{\lambda}^{+} {\hat T}_{\lambda
n}^{j}\right)+\Delta\nu_{j}{\hat T}_{nn}^{j}\right],
\label{eq18}\\
H_{N+1}^{(N)}=-\sum_{j=1}^{(N)} \tilde{\cal
H}_{j}^{(N)}=-\sum_{j=1}{\hat w}_{j}=
-A\sum_{k=1}^{N}\left[\sum_{\lambda=1}^{n-1}\varepsilon_{\lambda}
\sqrt{\Delta}(b_{\lambda}{\hat T}_{n\lambda}^{k}+b_{\lambda}^{+}
{\hat T}_{\lambda n}^{k})+\Delta\nu_{k}{\hat
T}_{nn}^{k}\right].\nonumber
\end{gather}
 They commute with $\big\{\tilde{\cal H}_{j}^{(N)}\big\}$
and ${\hat N}_{\alpha}=-\sum\limits_{j=1}^{N}{\hat
T}_{\alpha\alpha}^{j} +b_{\alpha}^{+}b_{\alpha}$.

 And f\/inally, let us choose as ${\hat T}_{\alpha\beta}^{j}$ the matrix
representation of $SL(n)$,
\[
{\hat T}_{\alpha\beta}^{j}\to{\hat R}_{\alpha\beta}^{j},
\]
 and add linear combination $\sum\limits_{\alpha=1}^{n-1}\omega_{\alpha}
{\hat N}_{\alpha}$ to $H_{N+1}^{(N)}.$

 It is easy to see that we obtain just the operator of generalized Dicke
 model \eqref{eq4}
 for $n\geq3$,
\[
H^{(N,n)}=\sum_{\alpha=1}^{n-1}\omega_{\alpha}\hat N_{\alpha}
+\sum_{k=1}^{N}\Delta_{k}\hat R_{nn}^{(k)}+\sum_{\alpha=1}^{n-1}
\sum_{k=1}^{N} g_{\alpha}^{(k)}\big[b_{\alpha}\hat
R_{n\alpha}^{(k)}+b_{\alpha}^{+} \hat R_{\alpha n}^{(k)}\big]
\]
 with coupling constants
\begin{gather}
g_{\alpha}^{(k)}=-A \varepsilon_{\alpha}\sqrt{\Delta}, \qquad
\varepsilon_{\alpha}=\pm1.\label{eq19}
\end{gather}
Hence, under the above choice of the coupling constants, the most
general model \eqref{eq4}  becomes quantum integrable and the
spectral problem might be simplif\/ied. However, we did not f\/ind
the way for doing it except the simplest nontrivial case of $N=2$
which is described in the next section.

\section[Explicit solution for $N=2$]{Explicit solution for $\boldsymbol{N=2}$}

In this case  the Hamiltonian \eqref{eq4} under the conditions
\eqref{eq19} can be written as
\begin{gather}
H_{n}^{(2)}=\sum_{\alpha=1}^{n-1}\omega_{\alpha} {\hat N}_{\alpha}
+{\Delta}_{1}{\hat R}_{nn}^{1} +\Delta_{2}{\hat R}_{nn}^{2}+
\sum_{\alpha=1}^{n-1}\sum_{k=1}^{2}\big[b_{\alpha}{\hat
R}_{n\alpha}^{k}
+b_{\alpha}^{+}{\hat R}_{\alpha n}^{k}\big],\label{eq20}\\
{\hat N}^{\alpha}=b_{\alpha}^{+}b_{\alpha}-\sum_{j=1}^2 {\hat
R}_{\alpha\alpha}^{j}.\nonumber
\end{gather}
 Let us rewrite \eqref{eq20} as
\begin{gather}
H_{n}^{(2)}=\sum_{\alpha=1}^{n-1}\omega_{\alpha}
{\hat N}_{\alpha}+{\hat k}_{1}+{\hat k}_{2},\label{eq21}\\
{\hat k}_{1,2}=\pm\left(\Delta_{1}-\Delta_{2}\right)^{-1} {\hat
P}_{1,2} +\sum_{\alpha=1}^{n-1}\big[b_{\alpha}{\hat
R}_{n\alpha}^{1,2}+b_{\alpha}^{+} {\hat R}_{\alpha
n}^{1,2}\big]+\Delta_{1,2}{\hat R}_{nn}^{1,2},\qquad {\hat
P}_{1,2}=\sum_{\alpha,\beta=1}^n {\hat R}_{\alpha\beta}^{1}{\hat
R}_{\beta\alpha}^{2}.\nonumber
\end{gather}
 Eigenfunctions of \eqref{eq21} are of course common eigenfunctions of $(n+1)$ operators
$\{N_{\alpha}\}$, ${\hat k}_{1}$, ${\hat k}_{2},$
\[
\psi=\left[\sum_{\lambda, \mu=1}^{n-1}A_{\lambda\mu}
\varphi_{\lambda}^{1}\varphi_{\mu}^{2}
b_{\lambda}^{+}b_{\mu}^{+}+\sum_{\lambda=1}^{n-1}(
B_{\lambda}^{(1)}\varphi_{\lambda}^{1}\varphi_{n}^{2}+B_{\lambda}^{(2)}
\varphi_{\lambda}^{2}\varphi_{n}^{1})b_{\lambda}^{+}+
C\varphi_{n}^{1}\varphi_{n}^{2}\right]Z_{n}| 0\rangle ,
\]
 where $\varphi_{\lambda}^{1}$, $\varphi_{\mu}^{2}$ are eigenvectors
of $\{{\hat R}_{\lambda\lambda}^{1}\}$, $\{{\hat
R}_{\mu\mu}^{2}\}$ and $ Z_{n}=\prod\limits_{\alpha=1}^{n-1}
\left(b_{\alpha}^{+}\right)^{N_{\alpha}}.$

One gets from the relations
 ${\hat k}_{1}\psi=k_{1}\psi$,
${\hat k}_{2}\psi=k_{2}\psi$ the following set of algebraic
equations for the coef\/f\/icients $\{A, B, C\}$,
\begin{gather*}
A_{\mu\lambda}(\Delta_{1}-
\Delta_{2})^{-1}+B_{\mu}^{(2)}=k_{1}A_{\lambda\mu},\\
B_{\lambda}^{(2)}(\Delta_{1}-\Delta_{2})^{-1}+C
=k_{1}B_{\lambda}^{(1)},\\
B_{\lambda}^{(1)}(\Delta_{1}-\Delta_{2})^{-1}+
\Delta_{1}B_{\lambda}^{(2)}+
\sum_{\mu}A_{\mu\lambda}(N_{\mu}+1+\delta_{\lambda\mu})
=k_{1}B_{\lambda}^{(2)},\\
 C(\Delta_{1}-\Delta_{2})^{-1}+C\Delta_{1}+\sum_{\mu}B_{\mu}^{(1)}
(N_{\mu}+1)=k_{1}C,\\
-A_{\mu\lambda}
(\Delta_{1}-\Delta_{2})^{-1}+B_{\lambda}^{(1)}=k_{2}A_{\lambda\mu},\\
-B_{\lambda}^{(1)}(\Delta_{1}-\Delta_{2})^{-1}
+C=k_{2}B_{\lambda}^{(2)},\\
-B_{\lambda}^{(2)}(\Delta_{1}-\Delta_{2})^{-1}+
\Delta_{2}B_{\lambda}^{(1)}+\sum_{\mu}
A_{\lambda\mu}(N_{\mu}+1+\delta_{\lambda\mu})=k_{2}B_{\lambda}^{(1)},\\
-C(\Delta_{1}-\Delta_{2})^{-1}+\Delta_{2}C+\sum_{\mu}
B_{\mu}^{(2)}(N_{\mu}+1)=k_{2}C.
\end{gather*}
 They can be easily reduced to two equations for $k_{1}$, $k_{2}$:
\begin{gather*}
(k_{1}+k_{2}-\Delta_{1})(\Delta_{1}-\Delta_{2})^{-1}+
(\Delta_{1}-\Delta_{2})^{-2}-k_{1}(k_{1}-\Delta_{1})+1
+\sum_{\mu=1}^{n-1}(N_{\mu}+1)=0,\\
(\Delta_{2}-k_{1}-k_{2})(\Delta_{1}-\Delta_{2})^{-1}+
(\Delta_{1}-\Delta_{2})^{-2}-k_{2}(k_{2}-\Delta_{2})+1+
\sum_{\mu=1}^{n-1}(N_{\mu}+1)=0,
\end{gather*}
 and the  eigenvalues of the Hamiltonian \eqref{eq20} are given by
\[
h^{2,n}=\sum_{\alpha=1}^{n-1}\omega_{\alpha}N_{\alpha}+k_{1}
+k_{2}.
\]

\section{Conclusions}
In this paper, we found the most general quantum integrable model
  of interaction
 of ($n-1)$ modes of radiation with $N$ dipole sources which comprises
all known models of that type:
   Jaynes--Cummings model ($N=1$, $n=2$), its generalisation for arbitrary
number of modes ($N=1$), Dicke model (arbitrary $N$, $n=2$). The
commutative ring of operators which includes $H_{n}^{(N)}$
 is found in explicit form. We did not use any ``anisotropic'' form
of the Gaudin solution; the Hamiltonian of the model was
 obtained
 via some limiting procedure. We almost immediately got
 solution for $N=2$, arbitrary $n$ with the use of additional
 integrals of motion, but the most interesting case
of solution for arbitrary $N$ and n  is unreachable at the present
stage of investigation. We hope to come back to this problem in
the future.

\bigskip

{\it This paper has been presented for memorial volume of Vadim
Kuznetsov. We both knew him personally~--- he was our guest in
Dubna almost 17 years ago when he was a~PhD student of Professor
I.V.~Komarov in Leningrad. We remember him as bright young man,
very active in the field of quantum and classical integrable
systems. Later he solved very complicated problem of integration
of equations of motion for classical Toda chains with
non-exponential ends~{\rm \cite{Kuznetsov}} proposed by one of us~(V.I.).}

\LastPageEnding

\end{document}